\documentclass[preprint,aps,12pt,preprintnumbers,nofootinbib,superscriptaddress]{revtex4}

\usepackage{graphicx}
\usepackage{amssymb,amsmath}
\usepackage{subfigure}
\def\lag{{\mathcal{L}}}

\def\A{{\mathbf{\tilde A}}}
\def\Dslash{D\hskip-0.65em /}
\DeclareMathOperator{\tr}{tr}

\begin{document}

%\newcount\hour \newcount\minute
%\hour=\time \divide \hour by 60
%\minute=\time
%\count99=\hour \multiply \count99 by -60 \advance \minute by \count99
%\newcommand{\mydate}{\ \today \ - \number\hour :00}

\preprint{CALT 68-2643}
\preprint{UCSD/PTH 07-04}

\title{The Lee-Wick Standard Model}
\author{Benjam\'in Grinstein}
\email[]{bgrinstein@ucsd.edu}
\affiliation{Department of Physics, University of California at San Diego, La Jolla, CA 92093}

\author{Donal O'Connell}
\email[]{donal@theory.caltech.edu}
\affiliation{California Institute of Technology, Pasadena, CA 91125}

\author{Mark B. Wise}
\email[]{wise@theory.caltech.edu}
\affiliation{California Institute of Technology, Pasadena, CA 91125}

\date{\today}

\begin{abstract}

We construct a modification of the standard model which stabilizes
the Higgs mass against quadratically divergent radiative corrections,
using ideas originally discussed by Lee and Wick in the context of a
finite theory of quantum electrodynamics. The Lagrangian includes new
higher derivative operators. We show that the higher derivative terms
can be eliminated by introducing a set of auxiliary fields; this allows
for convenient computation and makes the physical interpretation more
transparent. Although the theory is unitary, it does not satisfy the
usual analyticity conditions.

\end{abstract}

\maketitle

\section{Introduction}

The extreme fine-tuning needed to keep the Higgs mass small compared
to the Planck scale (i.e., the hierarchy puzzle) has motivated many
extensions of the minimal standard model. All of these contain new
physics, beyond that in the minimal standard model, which might be
observed at the Large Hadron Collider ({\rm LHC}). The most widely
explored of these extensions is low energy supersymmetry. In this paper
we introduce another extension of the standard model that solves the
hierarchy puzzle.

Our approach builds on the work of Lee and
Wick~\cite{Lee:1969fy,Lee:1970iw} who studied the possibility that
the regulator propagator in Pauli-Villars corresponds to a physical
degree of freedom. Quantum electrodynamics with a photon propagator that
includes the regulator term is a higher derivative version of QED. The
higher derivative propagator contains two poles, one corresponding to the
massless photon, and the other corresponding to a massive Lee-Wick-photon
(LW-photon). A problem with this approach is that the residue of the
massive LW-photon pole has the wrong sign. Lee and Wick argued that
one can make physical sense of such a theory. There is no problem
with unitarity since the massive LW-photon is not in the spectrum; it
decays through its couplings to ordinary fermions. However, the wrong
sign residue moves the poles in the photon two point function that are
associated with this massive resonance from the second sheet to the
physical sheet, introducing time dependence that grows exponentially.
Lee and Wick and Cutkosky \emph{et al.}~\cite{CLOP} propose a modification
of the usual integration contour in Feynman diagrams that removes
this growth and preserves unitarity of the S matrix\footnote{The
consistency of this approach is controversial~\cite{Nakanishi:1971jj}.}.  This was further
discussed in~\cite{Lee:1971ix,Coleman}.
%Altering the contour of energy
%integration in this way is a future boundary condition and so the theory,
%although unitary, has a modified causal structure. The effects of this
%modification, at low energies, are suppressed by powers of the mass of
%the LW-photon.

The theory of QED that Lee and Wick studied is finite. In this paper
we propose to extend their idea to the standard model, removing the
quadratic divergence associated with the Higgs mass, and thus solving the
hierarchy problem. In the LW-standard model, every field in the minimal
standard model has a higher derivative kinetic term that introduces a
corresponding massive LW-resonance. These masses are additional free
parameters in the theory and must be high enough to evade current
experimental constraints. For the non-Abelian gauge bosons the higher
derivative kinetic term has, because of gauge invariance, new higher
derivative interactions. Hence the resulting theory is not finite;
however, we argue that it does not give rise to a quadratic divergence
in the Higgs mass, and so solves the hierarchy puzzle. A power counting
argument and some explicit one loop calculations are given to demonstrate
this. For explicit calculations, and to make the physics clearer, it is
useful to remove the higher derivative terms in the Lagrangian density
by introducing auxiliary LW-fields that, when integrated out, reproduce
the higher derivative terms in the action.

The LW-standard model\footnote{LW extension of the standard model would
be more precise.} has a new parameter for each standard model field,
which corresponds physically to the tree-level mass of its LW-partner
resonance. Explicit calculations can be performed in this theory at
any order in perturbation theory, and the experimental consequences for
physics at the LHC, and elsewhere, can be studied. 
%Lee-Wick theories are
%unusual; for example, no consistent path integral formulation for the
%theory we propose is known~\cite{Boulware:1983vw}. 
The nonperturbative formulation of Lee-Wick theories has been studied in
~\cite{Kuti,Boulware:1983vw}.  Lee-Wick theories are unusual; however,
even if one does not take the particular model we present as the correct
theory of nature at the ${ \rm TeV}$ scale our work does suggest that
a further examination of higher derivative theories is warranted. Some
previous work on field theories with non-local actions that contain terms
with an infinite number of derivatives can be found in Ref.~\cite{nlocal}.

\section{A Toy Model}

To illustrate the physics of Lee-Wick
theory~\cite{Lee:1969fy,Lee:1970iw,Boulware:1983vw} in a simple setting,
we consider in this section a theory of one self-interacting scalar field,
$\hat \phi$, with a higher derivative term. The Lagrangian density is
\begin{equation}
\lag_\mathrm{hd} = \frac{1}{2} \partial_\mu \hat \phi \partial^\mu \hat \phi - \frac{1}{2 M^2} (\partial^2 \hat \phi)^2 - \frac{1}{2} m^2 \hat \phi^2 - \frac{1}{3!} g \hat \phi^3 ,
\label{eq:phiHDlag}
\end{equation}
so the propagator of $\hat \phi$ in momentum space is given by
\begin{equation}
\hat D(p) = \frac{i}{p^2 - p^4/M^2 - m^2} .
\end{equation}
For $M \gg m$, this propagator has poles at $p^2 \simeq m^2$ and also at
$p^2 \simeq M^2$. Thus, the propagator describes more than one degree
of freedom.

We can make these new degrees of freedom manifest in the Lagrangian
density in a simple way. First, let us introduce an auxiliary scalar field
$\tilde \phi$, so that we can write the theory as
\begin{equation}
\lag =  \frac{1}{2} \partial_\mu \hat \phi \partial^\mu \hat \phi  - \frac{1}{2} m^2 \hat \phi^2 - \tilde \phi \partial^2 \hat \phi + \frac{1}{2}M^2 \tilde \phi^2 - \frac{1}{3!} g \hat \phi^3 .
\label{eq:tmLagShift}
\end{equation}
Since $\lag$ is quadratic in $\tilde \phi$, the equations of motion of
$\tilde \phi$ are exact at the quantum level. Removing $\tilde \phi$
from $\lag$ with their equations of motion reproduces $\lag_\mathrm{hd}$
in Eq.~\eqref{eq:phiHDlag}.

Next, we define $\phi = \hat \phi + \tilde \phi$. In terms of this
variable, the Lagrangian in Eq.~\eqref{eq:tmLagShift} becomes, after
integrating by parts,
\begin{equation}
\lag = \frac{1}{2} \partial_\mu \phi \partial^\mu \phi - \frac{1}{2} \partial_\mu \tilde \phi \partial^\mu \tilde \phi + \frac{1}{2} M^2 \tilde \phi^2 - \frac{1}{2} m^2 ( \phi - \tilde \phi)^2 - \frac{1}{3!} g (\phi - \tilde \phi)^3.
\label{eq:scalarthy}
\end{equation}
In this form, it is clear that there are two kinds of scalar field:
a normal scalar field $\phi$ and a new field $\tilde \phi$, which we
will refer to as an LW-field. The sign of the quadratic Lagrangian of the
LW-field is opposite to the usual sign so one may worry about stability of
the theory, even at the classical level. We will return to this point. If
we neglect the mass $m$ for simplicity, the propagator of $\tilde \phi$
is given by
\begin{equation}
\tilde D(p) = \frac{-i}{p^2 - M^2} .
\end{equation}
The LW-field is associated with a non-positive definite norm
on the Hilbert space, as indicated by the unusual sign of its
propagator. Consequently, if this state were to be stable, unitarity of
the $S$ matrix would be violated. However, as emphasized by Lee and Wick,
unitarity is preserved provided that $\tilde \phi$ decays. This occurs
in the theory described by Eq.~\eqref{eq:scalarthy} because $\tilde \phi$
is heavy and can decay to two $\phi$-particles.

In the presence of the mass $m$, there is mixing between the scalar
field $\phi$ and the LW-scalar $\tilde \phi$. We can diagonalize this
mixing without spoiling the diagonal form of the derivative terms by
performing a symplectic rotation on the fields:
\begin{equation}
\begin{pmatrix} \phi \\ \tilde \phi \end{pmatrix}
=
\begin{pmatrix}
\cosh \theta && \sinh \theta \\
\sinh \theta && \cosh \theta 
\end{pmatrix}
\begin{pmatrix} \phi^\prime \\ \tilde \phi^\prime \end{pmatrix} .
\end{equation}
This transformation diagonalizes the Lagrangian if
\begin{equation}
\tanh 2 \theta = \frac{-2 m^2/M^2}{1- 2 m^2/M^2}.
\end{equation}
A solution for the angle $\theta$ exists provided $M > 2m$. The
Lagrangian~\eqref{eq:scalarthy} describing the system becomes
\begin{equation}
\lag = \frac{1}{2} \partial_\mu \phi^\prime \partial^\mu \phi^\prime - \frac{1}{2} m^{\prime2} \phi^{\prime2} 
- \frac{1}{2} \partial_\mu \tilde \phi^\prime \partial^\mu \tilde \phi^\prime 
+ \frac{1}{2} M^{\prime2} \tilde \phi^{\prime2} - \frac{1}{3!} (\cosh \theta - \sinh \theta)^3 g (\phi^\prime - \tilde \phi^\prime)^3 ,
\end{equation}
where $m^\prime$ and $M^\prime$ are the masses of the diagonalized
fields. Notice the form of the interaction; we can define $g^\prime =
(\cosh \theta - \sinh \theta)^3 g$ and then drop the primes to obtain
a convenient Lagrangian for computation.\footnote{In the following,
we will always assume that $M \gg m$ so that $g^\prime \simeq g$.}

Introducing the LW-fields makes the physics of the theory clear. There
are two fields; the heavy LW-scalar decays to the lighter scalar. At
loop level, the presence of the heavier scalar improves the convergence
of loop graphs at high energy consistent with our expectations from the
higher derivative form of the theory. We can use the familiar technology
of perturbative quantum field theory (appropriately modified~\cite{CLOP}) to compute quantum corrections to
the physics.

It is worth pausing for a moment to consider loop corrections to the two
point function of the LW-field. Using the one loop self energy, the 
full propagator for the LW-scalar is given, near $p^2 = M^2$, by
\begin{eqnarray}
\tilde D(p) &=&\frac{-i}{p^2 - M^2} + \frac{-i}{p^2 - M^2} (-i \Sigma(p^2)) \frac{-i}{p^2 - M^2} + \cdots \nonumber \\
&=& \frac{-i}{p^2 - M^2 + \Sigma(p^2)} .
\end{eqnarray}
Note that, unlike for ordinary scalars, there is a plus sign in front
of the self energy $\Sigma(p^2)$ in the denominator. 
%At one loop,
%the shift in the pole mass of the LW-scalar is $-\Sigma(M^2) / 2M$.
This sign is significant; for example, 
%from a one-loop computation we
%see that the imaginary part of the self energy is
%\begin{multline}
%{\rm Im }\Sigma(p^2)=-{g^2 \over {32 \pi} } \left( \theta(p^2- 4 m^2)\sqrt{1- \frac{4 m^2}{p^2}} + \theta(p^2 - 4 M^2) \sqrt{1-\frac{4 M^2}{p^2}} \right. \\
%\left. -2 \theta(p^2 - (M + m)^2) \sqrt{1 - 2 \frac{M^2 + m^2}{p^2} + \frac{(M^2 - m^2)^2}{p^4}} \right).
%\end{multline}
if one defines the width in the usual way (i.e., near the pole the
propagator has denominator $p^2-M^2+iM\Gamma$) then, from a one loop
computation of the self energy $\Sigma$, the width of the LW-field is (for Im $p^2 > 0$)
\begin{equation}
\Gamma = -\frac{g^2}{32 \pi M} \sqrt{1- \frac{4 m^2}{M^2}}.
\label{eq:LWwidth}
\end{equation}
This width differs in sign from widths of the usual particles we
encounter. With this result in hand, we can demonstrate how unitarity
of the theory is maintained in an explicit example. Consider $\phi\phi$
scattering in this theory. From unitarity, the imaginary part of the
forward scattering amplitude, $\mathcal M$, must be a positive quantity.
Near $p^2 = M^2$, the scattering is dominated by the $\tilde \phi$
pole and therefore the imaginary part of the amplitude is given for
Im $p^2 > 0$  by
\begin{equation}
\mathrm{Im } \mathcal{M} = - g^2 \frac{M \Gamma}{(p^2 - M^2)^2 + M^2 \Gamma^2}.
\end{equation}
The unusual sign of the propagator is compensated by the unusual sign
of the decay width.

As another consequence of this sign, the poles associated with these
LW-particles occur on the physical sheet of the analytic continuation
of the $S$ matrix, in violation of the usual rules of $S$ matrix
theory. These signs are also associated with exponential growth of
disturbances, which is related to the stability concerns alluded to
earlier. Lee and Wick, and Cutkowsky \emph{et al} argued that one can
nevertheless make sense of these theories by modifying the usual
contour prescription for momentum integrals. The Feynman $i \epsilon$
prescription can be thought of as a deformation of the contour such
that the poles on the real axis are appropriately above or below the
contour. 
%
%There
%is no issue in perturbation theory until one resums the widths into
%the propagators.
%
The Lee-Wick prescription is equivalent to imposing the boundary
condition that there are no outgoing exponentially growing modes.
It is well known that such future boundary conditions cause violations
of causality. In the Lee-Wick theory the acausal effects occur only on
microscopic scales, and show up as 
%peculiar behavior of resonances in
a peculiar time ordering of events; for example, the decay products of
a Lee-Wick particle appear at times before the Lee-Wick particle itself
is created.
%scattering experiments.
It is believed that this theory does not produce
violations of causality, or any paradoxes, on a macroscopic scale~\cite{Coleman}.

\newcommand\NOTHING[1]{{}}
\NOTHING
{\begin{figure}
\centering
\includegraphics[scale=0.8]{contour}
\put(-8,50){$p^0$}
\put(-9,49){\line(0,1){5}}
\put(-9,49){\line(1,0){5}}
\caption{The Lee-Wick prescription for the contour of integration in the
complex energy plane. The dots represent poles occurring in the integrand.}
\label{fig:LWcontour}
\end{figure}
}

\section{The Hierarchy Problem and Lee-Wick Theory}

In this section, we consider a scalar in the fundamental representation
interacting with gauge bosons. We find the Lagrange density for the LW
version of such a theory and show by power counting appropriate to the
higher derivative version of the theory that the scalar mass is free of
quadratic divergences. We then show by an explicit one loop calculation
that the ordinary scalar and the massive LW-fields do not receive a
quadratically divergent contribution to their pole masses.

\subsection{Gauge Fields}

The higher derivative Lagrangian in the gauge sector is 
\begin{equation}
\lag_\mathrm{hd} =  - \frac{1}{2} \tr \hat{F}_{\mu \nu} \hat{F}^{\mu\nu} + \frac{1}{M_A^2} 
\tr \left( \hat D^{\mu} \hat F_{\mu\nu} \right) \left( \hat D^\lambda \hat F_\lambda{}^\nu \right) ,
\label{eq:gbHDlag}
\end{equation}
where $\hat F_{\mu \nu} = \partial_\mu \hat A_\nu - \partial_\nu \hat
A_\mu - i g [ \hat A_\mu, \hat A_\nu]$, and $\hat A_\mu
= \hat A^A_\mu T^A$ with $T^A$ the generators of the gauge group $G$ in the
fundamental representation. We can now eliminate the higher derivative
term by introducing auxiliary massive gauge bosons $\tilde A$. Each
gauge boson is described by a Lagrangian
\begin{equation}
\lag = - \frac{1}{2} \tr \hat{F}_{\mu \nu} \hat{F}^{\mu\nu} - M_A^2 \tr \tilde A_\mu \tilde A^\mu + 2 \tr \hat F_{\mu\nu} \hat D^\mu \tilde A^\nu ,
\label{eq:gbLWlagNotDiag}
\end{equation}
where $\hat D_\mu \tilde A_\nu =
\partial_\mu \tilde A_\nu - i g [\hat A_\mu, \tilde A_\nu]$.
To diagonalize the kinetic terms, we introduce shifted fields defined by
\begin{equation}
\hat A_\mu = A_\mu + \tilde A_\mu.
\end{equation}
The Lagrangian becomes
\begin{multline}
\lag = - \frac{1}{2} \tr F_{\mu \nu} F^{\mu \nu} + \frac{1}{2} \tr \left( D_\mu \tilde A_\nu - D_\nu \tilde A_\mu \right)\left(D^\mu \tilde A^\nu - D^\nu \tilde A^\mu \right) 
-i g \tr \left( \left[ \tilde A_\mu, \tilde A_\nu \right] F^{\mu \nu} \right) 
\\
-\frac{3}{2} g^2 \tr \left( \left[ \tilde A_\mu, \tilde A_\nu \right]  \left[ \tilde A^\mu, \tilde A^\nu \right] \right) - 4 i g \tr \left( \left[ \tilde A_\mu, \tilde A_\nu \right] D^\mu \tilde A^\nu \right) - M_A^2 \tr \left(\tilde A_\mu \tilde A^\mu \right) .
\label{eq:gbLWlag}
\end{multline}
Note that for a $U(1)$ gauge boson all the commutators vanish, there
are no traces and an extra overall factor of $1/2$.

To perform perturbative calculations, we must introduce a gauge
fixing term. We could introduce such a term in the higher derivative
Lagrangian, Eq.~\eqref{eq:gbHDlag}, in terms of the Lagrangian
involving $A$ and $\tilde A$, Eq.~\eqref{eq:gbLWlag}, or even in
the Lagrangian with mixed kinetic terms for $\hat A$ and $\tilde A$,
Eq.~\eqref{eq:gbLWlagNotDiag}. As is usual in gauge theories, all of
these choices will yield the same results for physical quantities, but
they may differ for unphysical quantities. Different gauge choices can
differ on how divergent unphysical quantities are. Therefore, we will only
compute physical pole masses below.  In these computations, we introduce
a covariant gauge fixing term for the gauge bosons, $A_\mu^A$, in the
two field description of the theory given in Eq.~\eqref{eq:gbLWlag}. In
this choice of gauge, the propagator for the gauge bosons is given by
\begin{equation}
D_{\mu\nu}^{AB}(p) = \delta^{AB} \frac{i}{p^2} \left( \eta_{\mu\nu} -(1 - \xi) \frac{p_\mu p_\nu}{p^2} \right) ,
\label{eq:gbprop}
\end{equation}
while the propagator for the LW-gauge field is
\begin{equation}
\tilde D_{\mu\nu}^{AB}(p) = \delta^{AB} \frac{-i}{p^2 - M_A^2} \left( \eta_{\mu\nu} - \frac{p_\mu p_\nu}{M_A^2} \right).
\label{eq:LWprop}
\end{equation}

\vspace{10pt}

\subsection{Scalar Matter}

\vspace{10pt}

Let us move on to consider scalar matter transforming in the fundamental
representation of the gauge group. In ordinary field theory, such a
scalar field has a quadratic divergence in its pole mass.  The higher
derivative Lagrangian is given in terms of the scalar field $\hat \phi$ by
\begin{equation}
\lag_\mathrm{hd} = \left( \hat D_\mu \hat \phi \right)^\dagger \left(\hat D^\mu \hat \phi\right) - 
\frac{1}{M_\phi^2} \left( \hat D_\mu \hat D^\mu \hat \phi \right)^\dagger \left( \hat D_\nu \hat D^\nu \hat \phi \right) - V(\hat \phi).
\end{equation}
We eliminate the higher derivative term by introducing an LW-scalar multiplet $\tilde \phi$.
Then the Lagrangian is given in terms of the two fields $\hat \phi$ and $\tilde \phi$ by
\begin{equation}
\lag = \left( \hat D_\mu \hat \phi \right)^\dagger \left(\hat D^\mu \hat \phi\right) + M_\phi^2 \tilde \phi^\dagger \tilde \phi + \left( \hat D_\mu \hat \phi \right)^\dagger \left( \hat D^\mu \tilde \phi \right) + \left( \hat D^\mu \tilde \phi \right)^\dagger \left( \hat D^\mu \hat \phi \right) - V(\hat \phi),
\end{equation}
where the covariant derivative is
\begin{equation}
\hat D_\mu = \partial_\mu + i g \hat A^A_\mu T^A .
\end{equation}
For simplicity we take the ordinary scalar to have no potential at tree level, $V(\hat \phi) = 0$. 
%\begin{equation}
%V(\hat \phi) = \frac{\lambda}{4} \left( \hat \phi^\dagger \hat \phi \right)^2.
%\end{equation}
It is not hard to include a potential for $\hat \phi$ in the analysis,
and to show that the potential does not change our results.

We diagonalized the pure gauge sector by shifting the gauge fields;
in terms of the shifted gauge fields the hatted covariant derivative is
\begin{equation}
\hat D_\mu = D_\mu + i g \tilde A^A_\mu T^A,
\end{equation}
where $D_\mu = \partial_\mu + i g A^A_\mu T^A$ is the usual covariant derivative.
To diagonalize the scalar kinetic terms, we again shift the field
\begin{equation}
\hat \phi = \phi - \tilde \phi.
\end{equation}
The scalar Lagrangian becomes
\begin{eqnarray}
&&\lag = (D_\mu \phi )^\dagger D^\mu \phi - (D_\mu \tilde{\phi} )^\dagger D^\mu \tilde \phi + M_\phi^2 \tilde \phi^\dagger \tilde \phi 
+ i g (D^\mu \phi )^\dagger \tilde A^A_\mu T^A \phi 
+ g^2\phi^\dagger \tilde A^A_\mu T^A  \tilde A^{B \mu} T^B \phi \nonumber \\
&& - i g \phi^\dagger \tilde A^A_\mu T^A D^\mu \phi
- ig (D^\mu \tilde \phi )^\dagger \tilde A^A_\mu T^A \tilde \phi 
+ ig \tilde \phi^\dagger \tilde A^A_\mu T^A D^\mu \tilde \phi -g^2 \tilde \phi^\dagger \tilde A^A_\mu T^A  \tilde A^{B \mu} T^B \tilde \phi.
\end{eqnarray}
%where $V$ is given by the expression
%\begin{eqnarray}
%&&V(H, \tilde H) = \frac{\lambda}{4} \left( H^\dagger H \right)^2 + \frac{\lambda}{2} \left( H^\dagger H \right) \tilde H^\dagger \tilde H 
% - \frac{\lambda}{2} \left( H^\dagger H \right) \left(\tilde H^\dagger  H + H^\dagger \tilde H \right) 
%  \nonumber \\
%&&+ \frac{\lambda}{4} \left[ \left( H^\dagger \tilde H\right)^2 + \left( \tilde H^\dagger H\right)^2 + \left( \tilde H^\dagger \tilde H\right)^2 + 2 \left( H^\dagger \tilde H\right) \left( \tilde H^\dagger H\right) -2 \left( H^\dagger \tilde H\right) \left( \tilde H^\dagger \tilde H\right)\right. \nonumber \\ 
%&& \left.- 2 \left( \tilde H^\dagger H\right) \left( \tilde H^\dagger \tilde H\right) \right].
%\end{eqnarray}

\subsection{Power Counting}

Having defined the higher derivative and LW forms of the theory, we
present a power counting argument for the higher derivative version of
the theory which indicates that the only physical divergences in the
theory are logarithmic. Since the power counting argument depends on the
behaviour of Feynman graphs at high energies, we only need to consider
the terms in the Lagrangian which are most important at high energies.

%For simplicity we neglect the scalar potential (i.e., set $\lambda$ to zero ), however, it is not difficult to see that our conclusions also hold when $\lambda$ is non-zero.
%Schematically, the higher derivative Lagrangian density describing our theory is
%\begin{equation}
%\lag \sim - G^{\mu \nu} D^2 G_{\mu \nu} - H^\dagger D^4 H
%\end{equation}
For the perturbative power counting argument in the higher derivative
theory, it is necessary to fix the gauge. We choose to add a covariant
gauge fixing term $- (\partial_\mu \hat{A}^{A \mu})^2 / 2 \xi$ to the
Lagrange density and introduce Faddeev-Popov ghosts that couple to the
gauge bosons in the usual way. Then the propagator for the gauge field is
\begin{equation}
\hat D_{{\mu \nu}}^{AB}(p) = \delta^{AB}\frac{-i}{p^2 - p^4/M_A^2} \left( \eta_{\mu \nu} - (1 - \xi) \frac{p_\mu p_\nu}{p^2} - \xi \frac{p_\mu p_\nu}{M_A^2} \right) .
\end{equation}
We work in $\xi = 0$ gauge. Note that $\xi = 0$ corresponds to Landau
gauge and that the gauge boson propagator scales as $p^{-4}$ at high
energy. The propagator for the scalar in the fundamental representation is
\begin{equation}
\hat D^{ab}(p) = \delta^{ab}\frac{i}{p^2 - p^4/M_\phi^2 } .
\end{equation}
At large momenta the scalar propagator scales as $p^{-4}$ while the
Faddeev-Popov ghost propagator scales as $p^{-2}$, as usual. There are
three kinds of vertices: those where only gauge bosons interact, vertices
where gauge bosons interact with two scalars, and vertices where two
ghosts interact with one gauge boson. A vertex where $n$ vectors interact
(with no scalars) scales as $p^{6 -n}$ while a vertex with two scalars
and $n$ vectors scales as $p^{4 - n}$. The vertex between two ghosts
and one gauge field scales as one power of $p$, as usual.

Consider an arbitrary Feynman graph with $L$ loops, $I^\prime$ internal
vector lines, $I$ internal scalar lines, $I_g$ internal ghost lines,
and with $V_n^\prime$ or $V_n$ vertices with $n$ vectors and zero or two
scalar particles, respectively. We also suppose there are $V_g$ ghost
vertices. Then the superficial degree of divergence, $d$, is
\begin{equation}
d = 4 L - 4 I^\prime - 4 I - 2 I_g + \sum_n V_n^\prime (6-n) + \sum_n V_n (4-n) + V_g.
\end{equation}
We can simplify this expression using some identities. First, the number
of loops is related to the total number of propagators and vertices by
\begin{equation}
L = I + I^\prime +I_g - \sum_n (V^\prime_n + V_n) - V_g + 1 ,
\end{equation}
while the total number of lines entering or leaving the vertices is
related to the number of propagators and external lines by
\begin{equation}
\sum_n \left( n V^\prime_n + (n+2) V_n \right) + 3 V_g = 2 ( I+I^\prime +  I_g) + E + E^\prime + E_g,
\end{equation}
where $E$ is the number of external scalars, $E^\prime$ is the number of
external vectors, and $E_g$ is the number of external ghosts. Finally,
because the Lagrangian is quadratic in the number of scalars and ghosts,
the number of scalar lines and ghost lines is separately conserved. Thus,
\begin{equation}
2 \sum_n V_n = 2 I + E , \;\;\;\; 2 V_g = 2 I_g + E_g .
\end{equation}
With these identities in hand, we may express the superficial degree of
divergence as
\begin{equation}
d = 6 - 2 L -E - E^\prime - 2 E_g.
\end{equation}
Gauge invariance removes the potential quadratic divergence in the gauge
boson two point function. Scalar mass renormalizations have $E=2$, so that
$d = 4 - 2L$. Consequently, the only possible quadratic divergence in
the scalar mass is at one loop.  However, gauge invariance also removes
the divergence in the scalar mass renormalization, because two of the
derivatives must act on the external legs. To see this, note that the
interaction involves
\begin{equation}
\phi^\dagger D^{4} \phi \sim \phi^\dagger ( \partial^2  + \partial \cdot A + A \cdot \partial + A^2 )^2 \phi .
\end{equation}
Since we are working in Lorentz gauge, $\partial \cdot A = 0$. We
may ignore the $A^2$ term compared to the $A\cdot \partial$ term,
as it is less divergent. Thus the most divergent terms in the
interaction are $\phi^\dagger A\cdot \partial \partial^{2} \phi$
or $\phi^\dagger \partial^{2} A\cdot \partial \phi$, where the
$\phi$ acted on by the derivatives is an internal line. But by
integration by parts and use of the gauge condition, we see that,
at one loop, we can always take one of the derivatives to act on
the external scalar. Thus the theory at hand is at most logarithmically
divergent.\footnote{It may seem that adding operators with more
than four derivatives could yield a finite theory, but that is not
the case. These theories are still logarithmically divergent.}

\subsection{One loop Pole Mass}

The power counting argument above was presented in the higher
derivative version of the theory. As a check of the formalism we
show, in the LW version of the theory, that the shift in the pole
masses of the ordinary scalar, the LW-scalar and the LW-gauge boson
do not receive quadratically divergent contributions at one loop.
It is important to compute a physical quantity since it is for these
that the higher derivative theory and the theory with LW-fields
give equivalent results\footnote{We have fixed different gauges in
our discussion of the power counting argument in the higher derivative
theory and our explicit computations in the LW version of the theory.
Consequently, we can only expect agreement between these theories
for physical quantities.}. We perform the computations in Feynman
gauge, using the propagators in Eqs.~\eqref{eq:gbprop}
and~\eqref{eq:LWprop}, and regulate our diagrams where necessary
using dimensional regularization with dimension $n$.

\subsubsection{The normal scalar}

\begin{figure}
\centering
\includegraphics{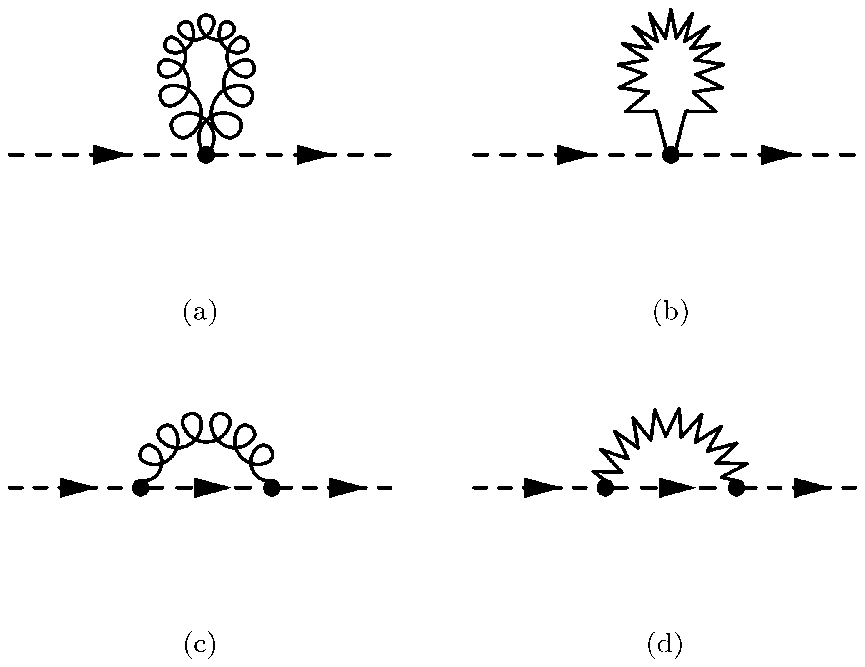}
\caption{One loop mass renormalization of the normal scalar field. The
curly line is a gauge field while the zigzag line is the LW-gauge
field. The dashed line represents the scalar field.}
\label{fig:normscalar}
\end{figure}

%\begin{fmffile}{setone}
%\begin{figure}[h]
%\centering
%\subfigure[]{
%\begin{fmfgraph*}(40,25)
%\fmfleft{i}
%\fmfright{o}
%\fmf{fermion,label=$p$}{i,v,o}
%\fmf{photon,tension=0.75}{v,v}
%\fmfdot{v}
%\end{fmfgraph*}
%}
%\hspace{0.1in}
%\subfigure[]{
%\begin{fmfchar*}(40,25)
%\fmfleft{i}
%\fmfright{o}
%\fmf{fermion,label=$p$}{i,v,o}
%\fmf{gluon}{v,v}
%\fmfdot{v}
%\end{fmfchar*}
%}
%\\
%\subfigure[]{
%\begin{fmfchar*}(40,25)
%\fmfleft{i}
%\fmfright{o}
%\fmf{fermion}{i,v,vp,o}
%\fmf{photon,left,tension=0}{v,vp}
%\fmfdot{v,vp}
%\end{fmfchar*}
%}
%\hspace{0.1in}
%\subfigure[]{
%\begin{fmfchar*}(40,25)
%\fmfleft{i}
%\fmfright{o}
%\fmf{fermion}{i,v,vp,o}
%\fmf{gluon,left,tension=0}{v,vp}
%\fmfdot{v,vp}
%\end{fmfchar*}
%}
%\caption{One loop mass renormalization of the normal scalar field.}
%\label{fig:normscalar}
%\end{figure}
%\end{fmffile}

At one loop, there are four graphs contributing to the scalar mass,
as shown in Figure~\ref{fig:normscalar}. We find
\begin{subequations}
\begin{align}
-i \Sigma_a(0) &= g^2 C_2(N) \int  \frac{d^n k}{(2 \pi)^n} \frac{n}{k^2}
\\
-i \Sigma_b(0) &= - g^2 C_2(N) \int \frac{d^n k}{(2 \pi)^n} \left( \frac{n-1}{k^2 - M_A^2} -\frac{1}{M_A^2} \right)
\\
-i \Sigma_c(0) &= -g^2 C_2(N) \int  \frac{d^n k}{(2 \pi)^n} \frac{1}{k^2} 
\\
-i \Sigma_d(0) &= - g^2 C_2(N)  \int  \frac{d^n k}{(2 \pi)^n} \frac{1}{M_A^2} .
\end{align}
\end{subequations}
We see that the quartic and quadratic divergences in this expressions
cancel in the sum, so that the mass is only logarithmically divergent.

\subsubsection{The LW-scalar}

\begin{figure}
\centering
\includegraphics{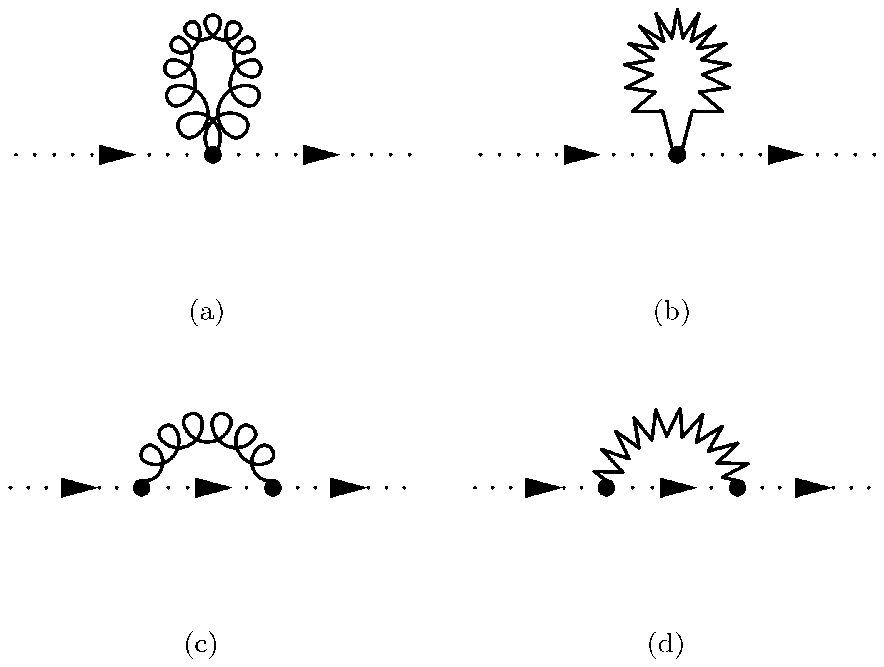}
\caption{One loop mass renormalization of the LW-scalar field. The dotted line represents the LW-scalar field while the other propagators are as in Figure~\ref{fig:normscalar}.}
\label{fig:abnormscalar}
\end{figure}

At one loop the shift in the pole mass is determined by the self energy
$\Sigma(p^2)$ evaluated at $p^2=M_\phi^2$.  The Feynman graphs are shown
in Figure~\ref{fig:abnormscalar}.
%\begin{fmffile}{settwo}
%\begin{figure}[h]
%\centering
%\subfigure[]{
%\begin{fmfchar*}(40,25)
%\fmfleft{i}
%\fmfright{o}
%\fmf{dashes_arrow}{i,v,o}
%\fmf{photon}{v,v}
%\fmfdot{v}
%\end{fmfchar*}
%}
%\hspace{0.1in}
%\subfigure[]{
%\begin{fmfchar*}(40,25)
%\fmfleft{i}
%\fmfright{o}
%\fmf{dashes_arrow}{i,v,o}
%\fmf{gluon}{v,v}
%\fmfdot{v}
%\end{fmfchar*}
%}
%\\
%\subfigure[]{
%\begin{fmfchar*}(40,25)
%\fmfleft{i}
%\fmfright{o}
%\fmf{dashes_arrow}{i,v,vp,o}
%\fmf{photon,left,tension=0}{v,vp}
%\fmfdot{v,vp}
%\end{fmfchar*}
%}
%\hspace{0.1in}
%\subfigure[]{
%\begin{fmfchar*}(40,25)
%\fmfleft{i}
%\fmfright{o}
%\fmf{dashes_arrow}{i,v,vp,o}
%\fmf{gluon,left,tension=0}{v,vp}
%\fmfdot{v,vp}
%\end{fmfchar*}
%}
%\caption{One loop mass renormalization of the abnormal scalar field.}
%\label{fig:abnormscalar}
%\end{figure}
%\end{fmffile}
We find
\begin{subequations}
\begin{align}
-i \Sigma_a(M_\phi^2) &= -g^2 C_2(N) \int  \frac{d^n k}{(2 \pi)^n} \frac{n}{k^2}
\\
-i \Sigma_b(M_\phi^2) &= g^2 C_2(N) \int \frac{d^n k}{(2 \pi)^n} \left( \frac{n-1}{k^2 - M_A^2} -\frac{1}{M_A^2} \right)
\\
-i \Sigma_c(M_\phi^2) &= g^2 C_2(N) \int  \frac{d^n k}{(2 \pi)^n} \left( \frac{1}{k^2 -2 p \cdot k} + \frac{4 M_\phi^2  - 4 p \cdot k}{k^2 ( k^2 - 2 p \cdot k)} \right)
\\
-i \Sigma_d(M_\phi^2) &= g^2 C_2(N)  \int  \frac{d^n k}{(2 \pi)^n} \left( \frac{1}{M_A^2} - \frac{4 M_\phi^2 - 2 p \cdot k}{(k^2 - M_A^2)(k^2 - 2 p \cdot k)} \right) .
\end{align}
\end{subequations}
Once again, the quartic and quadratic divergence cancel in the sum of
the graphs, so that there is only a logarithmic divergence in the mass
of the LW-scalar.

\subsubsection{The LW-vector}

\begin{figure}
\centering
\includegraphics{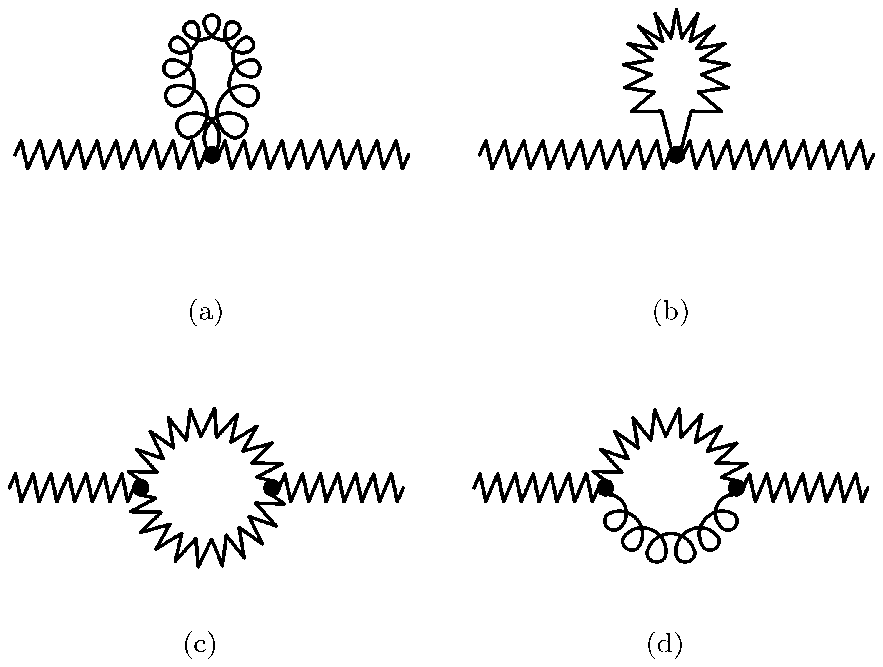}
\caption{One loop mass renormalization of the LW-vector field. The propagators are as in Figure~\ref{fig:normscalar}.}
\label{fig:abnormvector}
\end{figure}

For the LW-vectors the self energy tensor has the form
\begin{equation}
\Sigma^{AB}_{\mu \nu}(p^2)=\delta^{AB}\left[\Sigma(p^2)\eta_{\mu \nu}+\Sigma'(p^2)p_{\mu}p_{\nu}\right] .
\end{equation}
The shift in the pole mass is determined by $\Sigma(M_A^2)$. The
relevant graphs are shown in Figure~\ref{fig:abnormvector}. They are very
divergent. There are individual terms in Figure~\ref{fig:abnormvector}(c)
that diverge as the sixth power of a momentum cutoff. However these
cancel. There is also a quartic divergence in diagrams (b), (c) and
(d) that cancels between them. To check that the quadratic divergence
cancels we regulate the diagrams with dimensional regularization. In
$n$ dimensions, a quadratic divergence manifests itself as a pole at
$n=2$. Hence, we set $n=2-\epsilon$, expand about $\epsilon=0$ and
extract the $1/\epsilon$ part of $\Sigma(M_A^2)$.
%
%\begin{fmffile}{setthree}
%
%\begin{figure}[h]
%\centering
%\subfigure[]{
%\begin{fmfchar*}(40,25)
%\fmfleft{i}
%\fmfright{o}
%\fmf{gluon}{i,v,o}
%\fmf{photon}{v,v}
%\fmfdot{v}
%\end{fmfchar*}
%}
%\hspace{0.1in}
%\subfigure[]{
%\begin{fmfchar*}(40,25)
%\fmfleft{i}
%\fmfright{o}
%\fmf{gluon}{i,v,o}
%\fmf{gluon}{v,v}
%\fmfdot{v}
%\end{fmfchar*}
%}
%\\
%\subfigure[]{
%\begin{fmfchar*}(40,25)
%\fmfleft{i}
%\fmfright{o}
%\fmf{gluon}{i,v}
%\fmf{gluon,left,tension=0.5}{v,vp}
%\fmf{gluon,left,tension=0.5}{vp,v}
%\fmf{gluon}{vp,o}
%\fmfdot{v,vp}
%\end{fmfchar*}
%}
%\hspace{0.1in}
%\subfigure[]{
%\begin{fmfchar*}(40,25)
%\fmfleft{i}
%\fmfright{o}
%\fmf{gluon}{i,v}
%\fmf{gluon,left,tension=0.5}{v,vp}
%\fmf{photon,right,tension=0.5}{v,vp}
%\fmf{gluon}{vp,o}
%\fmfdot{v,vp}
%\end{fmfchar*}
%}
%\caption{One loop mass renormalization of the abnormal vector field.}
%\label{fig:abnormvector}
%\end{figure}
%\end{fmffile}
%
We find that
%
%\begin{equation}
%i \Sigma_a(M^2) = g^2 C_2(G) \int \frac{d^n k}{(2 \pi)^n} \frac{1}{k^2} (n-1)
%\end{equation}
%
%\begin{equation}
%i \Sigma_b(M^2) = - 3 g^2 C_2 (G) \int \frac{d^n k}{(2 \pi)^n} \frac{1}{k^2 -M^2} \left[ n - 1 - \frac{k^2}{M^2} \left(1 %- \frac{1}{n} \right) \right]
%\end{equation}
%
%In graphs (c) and (d) I include parts which are not just proportional to $\eta_{\mu \nu}$. 
%
%\begin{multline}
%i \Sigma_c(M^2) \eta_{\mu \nu} = 2 g^2 C_2(G) \int \frac{d^n k}{(2 \pi)^n} \left[ \frac{1}{k^2 - M^2} \eta_{\mu \nu} %\left(1 - \frac{k^2}{M^2} +  \frac{k^2}{n M^2} \right) \right.
%\\
%\left.
%+ \frac{1}{(k^2 -M^2)(k^2 -2 p \cdot k)} \left( \eta_{\mu \nu} \left( 3 k^2 + 4 M^2 - \frac{k^4}{M^2} \right) + k_\mu %k_\nu \left( 4n -8 + \frac{k^2}{M^2} \right) \right) \right]
%\end{multline}
%
%\begin{multline}
%i \Sigma_d(M^2) \eta_{\mu\nu} = -g^2 C_2(G) \int \frac{d^n k}{(2 \pi)^n} \left[ \frac{1}{k^2} \eta_{\mu \nu} 
%\right. 
%\\
%\left.
%+ \frac{1}{k^2 (k^2 - 2 p \cdot k)} \left( \eta_{\mu \nu} \left(3 k^2 + 4M^2 - \frac{k^4}{M^2} \right) + k_\mu k_\nu %\left(4n -8 + \frac{k^2}{M^2} \right) \right) \right]
%\end{multline}
\begin{subequations}
\begin{align}
-i \Sigma_a(M_A^2) &=  \frac{i g^2}{4 \pi} C_2(G) \left(-\frac{2}{\epsilon}\right)
\\
-i \Sigma_b(M_A^2) &= \frac{i g^2}{4 \pi} C_2(G) \left(\frac{3}{\epsilon} \right)
\\
-i \Sigma_c(M_A^2) &= \frac{i g^2}{4 \pi} C_2(G) \left(-\frac{6}{\epsilon}\right)
\\
-i \Sigma_d(M_A^2) &= \frac{i g^2}{4 \pi} C_2(G) \left(\frac{5}{\epsilon}\right) .
\end{align}
\end{subequations}
As expected, the $1/\epsilon$ pole cancels in the sum. Finally, we note
that there are quadratic divergences in $\Sigma^{AB}_{\mu\nu}(p^2)$. Only
the gauge invariant physical quantity $\Sigma(M_A^2)$ must be free of
quadratic divergences.

\section{Lee-Wick Standard Model Lagrangian}

Now that we have understood why the radiative correction to the Higgs
mass cancels in these higher derivative theories, we move on to discuss
the Lagrangian which describes the standard model extended to include
a Lee-Wick partner for each particle. The gauge sector is as before.

\subsection{The Higgs Sector}

A higher derivative Lee-Wick Higgs sector was considered previously
in~\cite{Kuti}.
We take the higher derivative Lagrangian for the Higgs doublet $\hat H$ to be
\begin{equation}
\lag_\mathrm{hd} = \left( \hat D_\mu \hat H \right)^\dagger \left(\hat D^\mu \hat H\right) - 
\frac{1}{M_H^2} \left( \hat D_\mu \hat D^\mu \hat H \right)^\dagger \left( \hat D_\nu \hat D^\nu \hat H \right) - V(\hat H),
\end{equation}
where the covariant derivative is given by
\begin{equation}
\hat D_\mu = \partial_\mu + i g \hat A^A_\mu T^A + i g_2 \hat W^a_\mu T^a + i g_1 \hat B_\mu Y,
\end{equation}
while the potential is
\begin{equation}
V(\hat H) = \frac{\lambda}{4} \left( \hat H^\dagger \hat H - \frac{v^2}{2} \right)^2.
\end{equation}
We can then eliminate the higher derivative term by introducing an
LW-Higgs doublet $\tilde H$. As before, we then diagonalize the
Lagrangian by introducing the shifted field $\hat H = H - \tilde H$.
To diagonalize the gauge field Lagrangian, we introduced Lee-Wick
gauge bosons $\tilde A$, $\tilde B$, and $\tilde W$ as well as the
usual gauge fields $A$, $B$ and $W$. In terms of these fields the
covariant derivative is
\begin{equation}
\hat D_\mu = D_\mu + i g \tilde A^A_\mu T^A + i g_2 \tilde W^a_\mu T^a + i g_1 \tilde B_\mu Y,
\end{equation}
where 
\begin{equation}
D_\mu = \partial_\mu + i g A^A_\mu T^A + i g_2 W^a_\mu T^a + i g_1 B_\mu Y
\end{equation}
is the usual standard model covariant derivative. We introduce the notation
\begin{equation}
\mathbf{\tilde A}_\mu = g \tilde A^A_\mu T^A + g_2 \tilde W^a_\mu T^a + g_1 \tilde B_\mu Y
\end{equation}
for the LW-gauge bosons.
%
%The Lagrangian is given in terms of the
%two doublets $\hat H$ and $\tilde H$ by
%\begin{equation}
%\lag = \left( \hat D_\mu \hat H \right)^\dagger \left(\hat D^\mu \hat H\right) + M_H^2 \tilde H^\dagger \tilde H + \left( \hat D_\mu \hat H \right)^\dagger \left( \hat D^\mu \tilde H \right) + \left( \hat D^\mu \tilde H \right)^\dagger \left( \hat D^\mu \hat H \right) - V(\hat H),
%\end{equation}
%where the covariant derivative is
%\begin{equation}
%\hat D_\mu = \partial_\mu + i g \hat A^A_\mu T^A + i g_2 \hat W^a_\mu T^a + i g_1 \hat B_\mu Y,
%\end{equation}
%while the potential is
%\begin{equation}
%V(\hat H) = \frac{\lambda}{4} \left( \hat H^\dagger \hat H - \frac{v^2}{2} \right)^2.
%\end{equation}
%We diagonalized the pure gauge sector by shifting the gauge fields;
%in terms of the shifted gauge fields the hatted covariant derivative is
%\begin{equation}
%\hat D_\mu = D_\mu + i g \tilde A^A_\mu T^A + i g_2 \tilde W^a_\mu T^a + i g_1 \tilde B_\mu Y,
%\end{equation}
%where $D_\mu$ is the usual standard model covariant derivative.
%
%To diagonalize the Higgs kinetic terms, we again shift the field
%\begin{equation}
%\hat H = H - \tilde H.
%\end{equation}
%Introducing
%\begin{equation}
%\mathbf{\tilde A}_\mu = g \tilde A^A_\mu T^A + g_2 \tilde W^a_\mu T^a + g_1 \tilde B_\mu Y,
%\end{equation}
%the Higgs Lagrangian becomes
The Lee-Wick form of the Higgs Lagrangian is then 
\begin{multline}
\lag = \left(D_\mu H \right)^\dagger D^\mu H - \left(D_\mu \tilde{H} \right)^\dagger D^\mu \tilde H + M_H^2 \tilde H^\dagger \tilde H - V(H, \tilde H)
+ i \left(D_\mu H \right)^\dagger \A^\mu H \\
- i H^\dagger \A_\mu D^\mu H + H^\dagger \A_\mu \A^\mu H 
- i\left(D_\mu \tilde H \right)^\dagger \A^\mu \tilde H + i \tilde H^\dagger \A_\mu D^\mu \tilde H - \tilde H^\dagger \A_\mu \A^\mu \tilde H ,
\end{multline}
where $V$ is given by the expression
\begin{eqnarray}
V(H, \tilde H) &=& V(H - \tilde{H}) \nonumber \\
&=& \frac{\lambda}{4} \left( H^\dagger H - \frac{v^2}{2} \right)^2 + \frac{\lambda}{2} \left( H^\dagger H - \frac{v^2}{2} \right) \tilde H^\dagger \tilde H - \frac{\lambda}{2} \left( H^\dagger H - \frac{v^2}{2} \right) \left(\tilde H^\dagger  H + H^\dagger \tilde H \right) 
\nonumber \\
& & + \frac{\lambda}{4} \left[ \left( H^\dagger \tilde H\right)^2 + \left( \tilde H^\dagger H\right)^2 + \left( \tilde H^\dagger \tilde H\right)^2 + 2 \left( H^\dagger \tilde H\right) \left( \tilde H^\dagger H\right) -2 \left( H^\dagger \tilde H\right) \left( \tilde H^\dagger \tilde H\right) \right. \nonumber \\
& & \left. - 2 \left( \tilde H^\dagger H\right) \left( \tilde H^\dagger \tilde H\right) \right].
\end{eqnarray}

In unitary gauge, we write
\begin{equation}
H = \begin{pmatrix}
0 \\ \frac{v+h}{\sqrt 2}
\end{pmatrix}, \;\;\;\;\;\;
\tilde H = \begin{pmatrix}
\tilde h^+ \\ \frac{\tilde h + i \tilde P}{\sqrt 2}
\end{pmatrix} .
\end{equation}
With this choice, the mass Lagrangian for the Higgs scalar, its partner, the
charged LW-Higgs and pseudoscalar LW-Higgs fields is
\begin{equation}
\lag_{\mathrm{mass}} = -\frac{\lambda}{4} v^2 ( h - \tilde h)^2 +\frac{M_H^2}{2} \left( \tilde h \tilde h + \tilde P \tilde P + 2 \tilde h^+ \tilde h^- \right).
\end{equation}
There is mixing between the usual Higgs scalar and its partner; this
mixing can be treated perturbatively. It is possible to diagonalize
the mass matrices of these particles via a symplectic rotation, which
preserves the diagonal form of the kinetic terms.

The Higgs vacuum expectation value induces masses for the gauge
bosons. First, we focus on the mass Lagrangian for the LW-gauge bosons. In
terms of the $SU(2)$ and $U(1)$ LW-gauge fields, the Lagrangian is
\begin{equation}
\lag_{\mathrm mass} = \frac{g_2^2 v^2}{8} \left( \tilde W^a_\mu \tilde W^{a \mu} \right) - \frac{g_1 g_2 v^2}{4} \tilde W^3_\mu \tilde B^\mu + \frac{g_1^2 v^2}{8} \tilde B_\mu \tilde B^\mu - \frac{M_1^2}{2} \tilde B_\mu \tilde B^\mu - \frac{M_2^2}{2} \tilde W^a_\mu \tilde W^{a\mu}.
\end{equation}
There is mixing between the $\tilde W^3$ and $\tilde B$ LW-gauge fields. We can diagonalize
this Lagrangian by writing
\begin{equation}
\begin{pmatrix}
\tilde W^3 \\ \tilde B
\end{pmatrix}
= 
\begin{pmatrix}
\cos \phi & \sin \phi \\
- \sin \phi & \cos \phi
\end{pmatrix}
\begin{pmatrix}
\tilde U \\ \tilde V
\end{pmatrix},
\end{equation} 
where the mixing angle is given by 
\begin{equation}
\tan 2 \phi = \frac{g_1 g_2 v^2}{2} \left( M_1^2 - M_2^2 + (g_2^2 - g_1^2) \frac{v^2}{4} \right)^{-1} .
\end{equation}
We expect that $M_{1,2}$ lie in the TeV range, so that $\phi$ is a small angle.

There is also mixing between the gauge fields and the LW-gauge fields. We will
treat this mixing perturbatively. The Lagrangian describing this mixing is
\begin{equation}
\lag_\mathrm{mix} = M_W^2 \left( W^+_\mu \tilde W^{- \mu} + \tilde W^+_\mu W^{- \mu} \right) + M_Z^2 Z_\mu \left( \cos \theta_W \tilde W^{3 \mu} - \sin \theta_W \tilde B^\mu \right), 
\end{equation}
where $\theta_W$ is the Weinberg angle and $M_W, \; M_Z$ are the usual tree level
standard model masses for the $W$ and $Z$ gauge bosons. One consequence of
the mixing is that there is a tree level correction to the electroweak $\rho$
parameter
\begin{equation}
\Delta \rho = \rho - 1 = - \frac{\sin^2 \theta_W M_Z^2}{M_1^2} .
\end{equation}
The current experimental constraint on this parameter is $|\Delta \rho|
\lesssim 10^{-3}$~\cite{pdg} which leads to $M_1 \gtrsim 1 \mathrm{ TeV}$.

\subsection{Fermion Kinetic Terms}

For simplicity, we discuss explicitly the case of a single left-handed quark doublet $Q_L$.
It is straightforward to generalize this work to the other representations, and to include
generation indices. 

The higher derivative theory is
\begin{equation}
\lag_\mathrm{hd} = \overline{\hat Q}_L i \hat \Dslash \,\hat Q_L + \frac{1}{M_Q^2} \overline{\hat Q}_L i \hat \Dslash \hat \Dslash \hat \Dslash \, \hat Q_L.
\label{eq:fermionHD}
\end{equation}
Naive power counting of the possible divergences in this higher derivative
theory shows that there are potential quadratic divergences in one loop
graphs containing two external gauge bosons and a fermionic loop. However,
gauge invariance forces these graphs to be proportional to two powers
of the external momentum so that the graphs are only logarithmically
divergent. In this case, this cancellation is most easily understood in
the LW description of the theory, which we now construct.

We eliminate the higher derivative term by introducing LW-quark doublets
$\tilde Q_L$, $\tilde Q_R^\prime$ which form a real representation of
the gauge groups. The Lagrangian in this formulation becomes
\begin{equation}
\lag = \overline{\hat Q}_L i \hat \Dslash \, \hat Q_L + M_Q \left( \overline{\tilde Q}_L \tilde Q_R^\prime + \overline{\tilde Q^\prime}_R \tilde Q_L \right)
+ \overline{\tilde Q}_L i \hat \Dslash \, \hat Q_L + \overline{\hat Q}_L i \hat \Dslash \, \tilde Q_L - \overline{\tilde Q^\prime}_R i \hat \Dslash \, \tilde Q^\prime_R .
\end{equation}
Eliminating the LW-fermions with their equations of motion
\begin{equation}
\tilde Q^\prime_R = - \frac{i \hat \Dslash}{M_Q} \hat Q_L, \;\;\; 
\tilde Q_L = \frac{\hat \Dslash \hat \Dslash}{M_Q^2} \hat Q_L ,
\end{equation}
reproduces the higher derivative Lagrangian, Eq.~\eqref{eq:fermionHD}.

To diagonalize the kinetic terms, we introduce the shift $\hat Q_L =
Q_L - \tilde Q_L$, and the Lagrangian becomes
\begin{eqnarray}
& & \lag = \overline{Q}_L i \Dslash \, Q_L - \overline{\tilde Q}_L i \Dslash \, \tilde Q_L - \overline{\tilde Q^\prime}_R i \Dslash \, \tilde Q_R^\prime
+ M_Q \left( \overline{\tilde Q}_L \tilde Q_R^\prime + \overline{\tilde Q^\prime}_R \tilde Q_L \right) \nonumber \\
& & - \overline{Q_L} \gamma_\mu \tilde{\mathbf{A}}^\mu Q_L
+ \overline{\tilde Q_L} \gamma_\mu \tilde{\mathbf{A}}^\mu \tilde Q_L + \overline{\tilde Q^\prime_R} \gamma_\mu \tilde{\mathbf{A}}^\mu \tilde Q_R^\prime .
\label{eq:fermionLWlag}
\end{eqnarray}
Note that $\tilde Q_L$ and $\tilde Q_R'$ combine into a single Dirac
spinor of mass $M_Q$.

\begin{figure}[t]
\centering
\includegraphics{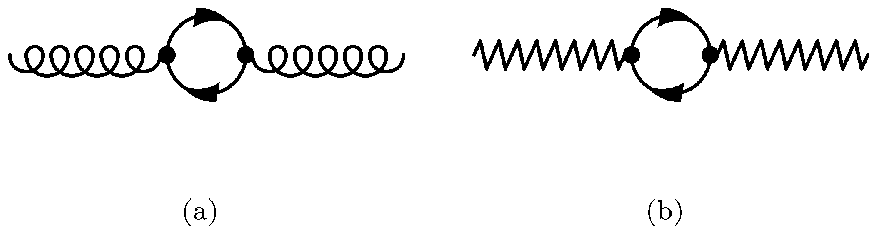}
\caption{One loop graphs involving fermions which are potentially
quadratically divergent. The solid lines represent fermion propagators
while the curly and zigzag lines represent gauge bosons and LW-gauge
bosons, respectively.}
\label{fig:fermionLoops}
\end{figure}

Now let us return to the issue of potential quadratic divergences in the
theory. Inspection of the Lagrangian, Eq.~\eqref{eq:fermionLWlag}, shows
that the only one loop graphs involving fermionic loops are the graphs
of Figure~\ref{fig:fermionLoops}. Figure~\ref{fig:fermionLoops}a is a
one loop correction to the gauge boson propagator, and consequently
is proportional to $p^2$, where $p$ is the momentum flowing into
the graph. Thus, the graph is logarithmically divergent, as is well
known. Figure~\ref{fig:fermionLoops}b is a one loop correction to
the LW-gauge boson propagator. One might think that this graph could
introduce a quadratic divergence of the LW-gauge boson mass. However,
the vertices between the fermions and the gauge bosons are equal to the
vertices between the fermions and the LW-gauge bosons, as can be seen
in Eq.~\eqref{eq:fermionLWlag}. Thus, Figure~\ref{fig:fermionLoops}b is
logarithmically divergent. Higher loop graphs in the theory are at most
logarithmically divergent by power counting.

\subsection{Fermion Yukawa Interactions}

To simplify the discussion in this section, we will neglect neutrino
masses. In the higher derivative formulation, the fermion Yukawas are
\begin{equation}
\lag_Y = g_u^{ij} \overline{\hat u^i}_R \hat H \epsilon \hat Q_L^j - g_d^{ij} \overline{\hat d^i}_R \hat H^\dagger \hat Q_L^j - g_e^{ij} \overline{\hat e^i}_R \hat H^\dagger \hat L_L^j + \mathrm{ h.c.} ,
\label{eq:HDyukawa}
\end{equation}
where repeated flavor indices are summed. In the formulation of the
theory in which there are no higher derivatives, and in which the kinetic
terms are diagonal, this becomes
\begin{eqnarray}
&& \lag_Y = g_u^{ij} (\overline{u^i}_R - \overline{\tilde u^i}_R) (H - \tilde H) \epsilon (Q_L^j - \tilde Q_L^j) - g_d^{ij} (\overline{d^i}_R - \overline{\tilde d^i}_R ) (H^\dagger - \tilde H^\dagger) (Q_L^j - \tilde Q_L^j) \nonumber \\
&& - g_e^{ij} (\overline{e^i}_R - \overline{\tilde e^i}_R) (H^\dagger - \tilde H^\dagger) (L_L^j - \tilde L_L^j) + \mathrm{ h.c.} .
\label{eq:LWyukawa}
\end{eqnarray}
The presence of the LW-fields in this equation improves the degree of
convergence at one loop. For example, consider a one loop correction
to the Higgs two point function coming from the first term of
Eq.~\eqref{eq:LWyukawa}. Various degrees of freedom can propagate
in the loop: the $u_R$ and $Q_L$ quarks, and also the $\tilde u_R$
and $\tilde Q_L$ LW-quarks. The presence of the LW-quarks cancels
the quadratic divergence in the loop with only the quarks. The sum of
these four graphs reproduces the result one would find by computing the
corresponding correction in the higher derivative formulation of the
theory, Eq.~\eqref{eq:HDyukawa}.

To simplify the flavor structure of the theory, we adopt the principle
of minimal flavor violation~\cite{MFV}. This forces all LW-fermions in the same
representation of the gauge group have the same mass. Now the Yukawas
can be diagonalized in the standard fashion. For notational brevity,
we choose to use the same symbol for the weak and mass eigenstates. In
terms of the mass eigenstate fields\footnote{They are mass eigenstate
fields when mixing between the normal and LW-fields is neglected. This
mixing can be treated as a perturbation.},
\begin{eqnarray}
&& \lag_Y = \frac{\sqrt 2}{v} \sum_i \left[ m_u^i (\overline{u^i}_R - \overline{\tilde u^i}_R) (H - \tilde H) \epsilon (Q_L^i - \tilde Q_L^i) - m_d^i (\overline{d^i}_R - \overline{\tilde d^i}_R ) (H^\dagger - \tilde H^\dagger) (Q_L^i - \tilde Q_L^i) \right. \nonumber \\
&& \left.
- m_e^i (\overline{e^i}_R - \overline{\tilde e^i}_R) (H^\dagger - \tilde H^\dagger) (L_L^i - \tilde L_L^i) + \mathrm{ h.c.} \right] ,
\end{eqnarray}
where 
\begin{equation}
Q_L = \begin{pmatrix}
u_L \\ V d_L
\end{pmatrix}, \;\;\;
\tilde Q_L = \begin{pmatrix} 
\tilde u_L \\ V \tilde d_L
\end{pmatrix}, \;\;\;
\tilde Q_R^{\prime} = \begin{pmatrix}
\tilde u_R^\prime \\ V\tilde d_R^\prime
\end{pmatrix}.
\end{equation}
Here $V$ is the usual CKM matrix. The LW-fermions decay via the Yukawa
interactions; for example, $\tilde \nu_e \to e^- \tilde h^+
\to e^- t \bar b$. LW-gauge bosons can decay to pairs of ordinary
fermions. All the heavy LW-particles decay in this theory, so the only
sources of missing energy in collider experiments are the usual standard
model neutrinos.

\section{Conclusions}

In this paper we have developed an extension of the minimal standard model
that is free of quadratic divergences. It is based on the work of Lee and
Wick who constructed a finite version of QED by associating the regulator
propagator in Pauli-Villars with a physical degree of freedom. Our model
is a higher derivative theory and as such contains propagators with wrong
sign residues about the new poles. Lee and Wick, and Cutkosky \emph{et
al.}  provide a prescription for handling this issue. The LW-particles
associated with these new poles are not in the spectrum, but instead
decay to ordinary degrees of freedom. Their resummed propagators do
not satisfy the usual analyticity properties since the poles are on the
physical sheet. Lee and Wick (see also Cutkosky \emph{et al.}) propose
deforming integration contours in Feynman diagrams so that there is no
catastrophic exponential growth as time increases. This amounts to a
future boundary condition and so LW-theories violate the usual causal
conditions. While the Lee Wick interpretation is peculiar it seems to
be consistent, at least in perturbation theory, and predictions for
physical observables can be made order by order in perturbation theory.

Since the extension of the standard model presented here is free of
quadratic divergences it solves the hierarchy problem. Our theory contains
one new parameter, the mass of the LW-partner, for each field. We reduced
the number of parameters by imposing minimal flavor violation to simplify
the flavor structure of the theory. To make the physical interpretation
clearer and the calculations easier we introduced auxiliary LW-fields. The
Lagrangian written in terms of these fields does not contain any higher
derivative terms. When the LW-fields are integrated out the higher
derivative theory is recovered.

This paper focused on the the structure of the Lagrange density for the
Lee-Wick extension of the standard model. We constructed the Lagrange
density, examined the divergence structure and showed how to introduce
auxiliary fields to clarify the physical interpretation. For the future,
a more extensive discussion of the phenomenology of the theory including
its implications for LHC physics is appropriate.

\acknowledgements

DOC would like to thank Stephen Adler for a helpful discussion and for
pointing out a useful reference. The work of BG was supported in part
by the US Department of Energy under contract DE-FG03-97ER40546, while
the work of DOC and MBW was supported in part by the US Department of
Energy under contract DE-FG03-92ER40701.

\end{document}